\documentclass[12pt,a4paper]{article}
\usepackage{listings}
\usepackage{color}
\usepackage{amsmath}
\usepackage{amssymb}
\usepackage{hyperref}
\usepackage{braket}
\usepackage{indentfirst}
\usepackage{graphicx}
\usepackage{bm}
\usepackage{epstopdf}
\usepackage{lmodern}
\usepackage[T1]{fontenc}
\usepackage{authblk}
\usepackage[margin=0.5in]{geometry}
\usepackage[square,sort,comma,numbers]{natbib}
\usepackage[font=scriptsize]{subcaption}
\usepackage{multirow}
\usepackage{xcolor,colortbl}

\definecolor{LightGray}{gray}{0.85}

\title{Implementation of a Quantum Algorithm to Estimate the Energy of a Particle in a Finite Square Well Potential on IBM Quantum Computer}
\author{Sina Shokri}
\author{Shahnoosh Rafibakhsh \thanks{Corresponding author: rafibakhsh@ut.ac.ir}}
\author{Roghayeh Pooshgan}
\author{Rita Faeghi}
\affil{Plasma Physics Research Center, Science and Research Branch, Islamic Azad University, Tehran, Iran}
\date{\today}                    

\begin{document}
  \maketitle
\section{Abstract}
In this paper, we implement a quantum algorithm -on IBM quantum devices, IBM QASM simulator and PPRC computer cluster -to find the energy values of the ground state and the first excited state of a particle in a finite square-well potential. We use the quantum phase estimation technique and the iterative one to execute the program on PPRC cluster and IBM devices, respectively. Our results obtained from executing the quantum circuits on the IBM classical devices show that our circuits
succeed at simulating the system. However, duo to scattered results, we execute only the iterative phase estimation part of the circuit on the 5 qubit quantum devices to reduce the circuit size and obtain low-scattered results. 

\section{Introduction}\label{sec:intro}
Simulation of quantum mechanics is considered as a difficult task especially for large systems. One of the main problems is the large volume of computer memory needed to store the quantum state of a physical system. The quantum state of a system with the computational basis $ \ket{x_1,x_2,...,x_n}$ is determined by $2^n$ amplitudes which becomes too large when $n$ grows. A suggested solution to this problem proposed by Richard Feynman and Yuri Manin \cite{feynman,manin}. They realized that a quantum computer had the potential to simulate quantum systems more efficiently than a classical one, and today quantum computing has become one of the most
growing fields in physics. During the last years, there have been many reasons to develop quantum computers, that two of them are significant. Firstly, applications of quantum simulation have proliferated in physics, chemistry, biology and other scientific fields \cite{lidar,guzik,lanyon,feng,kassal,savel,Reiher,YAMAGUCHI,daniel,Rakhmanov,walter,barends,Jordan1130,byrnes,markus,felice}. Secondly, the required technology to control quantum systems has improved, which enables us to perform more accurate simulations.

Since 2016, IBM has introduced a Cloud-based platform called IBM Q Experience \cite{ibmqe} which allows users to access a set of prototype quantum processors (QPUs) and execute algorithms and experiments on them \cite{alsina,Wang_2018,devitt,Beheraa_2019,zulehner2018,Behera_2019,Manabputra_2020,Joy_2019,Swain_2019,Maitra_2020,bochkin2020,kole}. This platform currently consists of six 5 qubit devices, one 16 qubit and one 1 qubit devices together with a simulator used for checking algorithms before running real experiments. The physical realization of these devices is on the basis of transmon qubits. There are basically two ways to write and run algorithms on the devices; Quantum Composer, which is an online GUI, allowing users to construct quantum algorithms by constructing quantum circuits. However, this approach can only be used to run quantum algorithms on the 5 qubit devices and is more suitable for small circuits. Alternatively, users may design circuits, in the scripting mode, by means of the QASM language, a language for creating quantum circuits invented by the IBM Q Experience team. The second means, on the other hand, is by writing python codes and running them using a python software developing kit (SDK) named QISKit \cite{qiskit_org}, and is appropriate for all types of algorithms. Hereby, the work we demonstrate in this paper is carried out using QISKit. 

The quantum devices that are publicly accessible over the Cloud are denoted by \textsl{IBM Q 5 Yorktown (ibmqx2)}, \textsl{IBM Q Burlington}, \textsl{IBM Q 5 London}, \textsl{IBM Q Essex}, \textsl{IBM Q Vigo} and \textsl{IBM Q Ourense} -the six  5 qubit devices- as well as \textsl{IBM Q 16 Melbourne} and \textsl{IBM Q Armonk} -the 16 qubit and 1 qubit devices, respectively. The classical backend used for simulation is named IBMQ QASM simulator. All backends work with a set of quantum gates composed of single-qubit rotational and phase-shift gates. All other single-qubit gates (such as $X, S, R_z$ etc.) are constructed, in general, by sequences of these three gates, which together with the CNOT, form the universal set of quantum gates. In addition to the number of qubits, the quantum devices mentioned are also different in terms of the qubit connectivity or topology, which the IBM Q Experience refers to as the \textit{coupling map} of the devices \cite{ibmq_device_spec}. 

In this paper, we modify and implement the quantum algorithm studied in Ref.~\cite{nakao} on IBM quantum computer to find the energy eigenvalues of the ground- and first-excited states of the one-dimensional Schrodinger equation of the finite square-well potential using the phase estimation technique. We use a trial wave function as the initial state and discretize it in the position and momentum spaces. We also construct the time-evolution matrix in the Hilbert space in which the computational basis vectors (i.e. qubit states) are defined. Then, we apply the time-evolution circuit to an initially prepared register and the phase estimation method is used to obtain the phase which includes the energy. We show that the proposed algorithms can achieve the desired results with reasonable errors. We discuss the implementation of the iterative phase estimation method in addition to the well-known quantum phase estimation scheme to reduce the size of the circuit and the number of qubits to effectively use the IBM quantum computing resources. Most importantly, to achieve the most out of the 5-qubit IBM backends, we shorten the circuit size -from 8 qubits used in Ref.~\cite{nakao} to 5 - by opting for the iterative phase estimation technique.

This paper is organized as follows. Section \ref{sec:algorithm} describes the steps of the quantum algorithm based on the phase estimation method. To perform a Digital Quantum Simulation, we need to engineer the time evolution operator to find the energy eigenvalues of the system. Moreover, the coordinate should be discretized and the initial wave function is approximated on the mesh points. We also explain two phase estimation algorithms used in this paper. In section \ref{sec:circop}, we explain how to construct quantum gates for the kinetic and potential terms in the time evolution operator. Results and discussion are given in section \ref{sec:res} and in section \ref{sec:conclusion} final remarks are discussed.

\section{Steps of the Quantum Algorithm}\label{sec:algorithm}
In this section, we present the steps of the quantum algorithm used in this paper, based on the phase estimation technique. 

If $\hat U$ is a unitary operator, the eigenvalue of such an operator is $e^{2 \pi i \theta}$ with $0\leq \theta <1$. The phase estimation technique aims to estimate the eigenvalues of the unitary operator and then $\theta$ as accurately as possible. In our work, $\hat U=\exp(-i\hat H t)$ is the time evolution operator with the eigenvalue $e^{-iE_nt}$. Therefore, the energy $E_n$ is given by
\begin{equation}
E_n=\frac{-2\pi \theta}{t}.
\end{equation}\label{eq:E}
We use Digital Quantum Simulation (DQS) to find the energy phase. When a real system is simulated in a quantum computer using DQS, we usually seek a simulating Hamiltonian $H^\prime$, possibly having deviation from the exact Hamiltonian $H$ \cite{zalka}. Then, we construct the time-evolution $U^\prime(t)= \exp(-iH^\prime t)$ and prepare the trial initial state $\ket{\phi(0)}$ - again possibly deviated from the exact initial state $\ket{\psi(0)}$ - in the quantum register to simulate the time-independent Schrodinger equation and eventually to find the energy spectrum through the phase estimation. In the following sections, we engineer $U^\prime(t)$ -and hence $H^\prime$- through a sequence of quantum gates by discretizing the position and momentum spaces.

\subsection{Time evolution}\label{sec:time}
To construct the time-evolution gate and in order to be engineered through the available quantum gates, the Hamiltonian should be written in terms of the kinetic operator $\hat K=\frac{\hat P^2}{2}$ and the potential operator $V(\hat X)$. Throughout this paper, we set $m=1$. The time interval $t$ is also divided into $n$ steps, $t=n\Delta t$. As the operators $\hat P$ and $V(\hat X)$, in the Hamiltonian, do not necessarily commute, the second-order Trotter formula is used to calculate the time evolution operator of each step \cite{trotter}: 
\begin{eqnarray}\label{eq:trotter}
\hat U(\Delta t)&=&\exp(-i(\hat K+V(\hat X))\Delta t)\\\nonumber
&=&\exp(-iV(\hat X)\frac{\Delta t}{2})\exp(-i\hat K{\Delta t})\exp(-iV(\hat X)\frac{\Delta t}{2})+O ((\Delta t)^3).
\end{eqnarray}
This expansion is used, as it will be shown, so that we can transform the time-evolution operator into elementary quantum gates. Now, the time-evolution operator should be applied to the initial state $\ket{\psi(0)}$ at $t=0$ and, in the coordinate space, it is written as the following:
\begin{equation}\label{eq:x-space}
\bra x \hat U(t) \ket{\psi(0)}=\bra x\exp(-iV(\hat X)\frac{\Delta t}{2})\exp(-i\hat K{\Delta t})\exp(-iV(\hat X)\frac{\Delta t}{2})\ket{\psi(0)}.
\end{equation}
As seen in Eq.~\eqref{eq:x-space}, $\exp(-i\hat K{\Delta t})$ is an operator in the momentum space and the Fourier transformation of the wave function is required:
\begin{eqnarray}\label{eq:FT}
\braket{p|\psi(0)}&=&\int_{-\infty}^{+\infty}e^{-2\pi i px} \psi(x,0) dx=U_{FT}^\dagger \psi(x,0),\\ \nonumber
\braket{x|\psi(0)}&=&\int_{-\infty}^{+\infty}e^{+2\pi i px} \psi(p,0) dx=U_{FT} \psi(p,0) .
\end{eqnarray}
Using the above equations, Eq.~\eqref{eq:x-space} could be written as the following:
\begin{eqnarray}\label{eq:x-pspace}
\bra x \hat U(t) \ket{\psi(0)}=\exp(-iV(x)\frac{\Delta t}{2})U_{FT}\exp(-i\hat K{\Delta t})U_{FT}^\dagger\exp(-iV(x^\prime)\frac{\Delta t}{2})\psi(x^\prime,0).
\end{eqnarray}
Now, we discretize the coordinate and explain how to approximate the wave function on the mesh points.

\subsection{Coordinate Discretization and QFT}\label{sec:xdis}
Consider a finite region $-d < x < d$ in the $x$-space. For an n-qubit register, we have $N = 2^n$ computational basis vectors in the set \{$\ket{k}$\}, where $k = 0,1, ... , N - 1$. Now, if we divide this region into $2^n$ intervals, and let each basis vector $\ket{k}$ represent a point within or on the boundaries of each interval, then the wave function $\psi(x)$ can be approximately expressed by the sum \cite{zalka,nakao,beneti}

\begin{equation}\label{eq:state1} 
\ket{\psi} = \sum_{k=0}^{N-1} \psi(x_k) \ket{k}.
\end{equation}
In other words, the basis vectors $\ket{k}$ represent the points on a one-dimensional grid with intervals $\frac{1}{N}$ between each pair of points. Now, in our discrete $x$-space spanned by the computational basis vectors $\ket{k}$, the probability $P_k$ of a basis ket $\ket{k}$ must be equal to the probability of the finding the particle in the region $\Delta x = \frac{1}{N}$ about the point $x_k$:

\begin{equation}\label{eq:pk}
P_k =  \int_{x_k-\frac{\Delta x}{2}}^{x_k+\frac{\Delta x}{2}} |\psi(x)|^2 dx.
\end{equation}

We resort to the symmetric distribution about $x=0$ for two reasons: Firstly, we take advantage of the convenience and the appealing appearance of such a distribution. Secondly, this procedure is a benchmark for other potentials, e.g. Coulomb potential, which are problematic at $x=0$. Furthermore, to make sure, for this work we implemented the asymmetric distribution (not discussed in this paper) with almost no changes in the final results. Such a symmetric distribution of points in the region $-d < x < d$ is given by the equation

\begin{equation}\label{eq:xdis}
x_k = -d + \left(\frac{k}{N} + \frac{1}{2N}\right). 
\end{equation}

Now, to write Eqs.~\eqref{eq:FT} in the discrete form, we need to use Quantum Fourier Transform (QFT). The QFT on a computational orthonormal basis vectors $\ket{0},\ket{1}, ... ,\ket{N-1}$ is defined as a unitary operation whose action on the basis vectors is as follows:
\begin{equation}\label{eq:qft}
\ket{j} \longrightarrow \frac{1}{\sqrt{N}}\sum_{k=0}^{N-1} e^{i2\pi jk/N} \ket{k}.
\end{equation}
In order to understand this, let us see what happens to the Fourier transformations of Eqs.~\eqref{eq:FT} when the coordinate and thus the momentum becomes discrete. That is, the Discrete Fourier Transformations (DFT) of $\psi(x)$ and $\phi(p)$:

\begin{eqnarray}\label{eq:qftinft}
\phi(p)& =&\frac{1}{\sqrt{N}}\sum_{x_k=-d}^{d} e^{-ip_jx_k} \psi(x_k) = U^{\dagger}_{DFT}\psi(x), \\\nonumber
\psi(x)& =&\frac{1}{\sqrt{N}}\sum_{p_j=-q}^{q} e^{ip_jx_k} \phi(p_j) = U_{DFT}\phi(p),
\end{eqnarray}
where $p_j$ represents the discretized momentum (i.e a one-dimensional grid in the $p$-space) whose corresponding distribution in the yet unknown range $-q < p < q$ can be found from the boundary condition that we impose on the wave function \cite{zalka}:
\begin{equation}\label{eq:ftbc}
\psi(x_k+2d) = \psi(x_k).
\end{equation}

For a symmetric distribution about the point $p=0$ in the $p$-space, we obtain
\begin{equation}\label{eq:pdis}
p_j = \frac{2\pi}{2d}\left(j + \frac{1}{2} - \frac{N}{2}\right) \quad , \quad j = 0,1, ... ,N-1. 
\end{equation}
Now, now we have a distribution of momentum, similar to that of coordinate in Eq.~\eqref{eq:xdis}, in the region $-q < p < q$ where $q = 2\pi N/4d$ and the particle's wave function in the momentum space can be approximately written as 

\begin{equation} 
\ket{\phi} = \sum_{j=0}^{N-1} \phi(p_j) \ket{j}
\end{equation}
The periodic boundary condition for the particle's wave function in the momentum space becomes 
\begin{equation}
\phi(p_j+2q) = \phi(p_j).
\end{equation} 

Finally, the Fourier transformation of Eq.~\eqref{eq:qftinft} applied to the state $\ket{\phi}$ of our quantum register gives
\begin{equation} 
U_{DFT}\ket{\phi} = \sum_{k=0}^{N-1} \psi(x_k) \ket{k}.
\end{equation}
That is, we have defined a unitary operator, $U_{DFT}$, which transforms a basis state $\ket{j}$ as follows
\begin{equation}\label{eq:uft}
\ket{j} \longrightarrow \frac{1}{\sqrt{N}}\sum_{k=0}^{N-1} e^{ip_jx_k} \ket{k}.
\end{equation}
This is not exactly the same as how QFT acts on a basis state as in Eq.~\eqref{eq:qft}. Nevertheless, substituting $p_j$ and $x_k$ from Eqs.~\eqref{eq:xdis} and \eqref{eq:pdis} respectively, we get
\begin{equation}\label{eq:uft1}
\ket{j} \longrightarrow \frac{1}{\sqrt{N}}\sum_{k=0}^{N-1} e^{i2\pi(j - N/2 + 1/2)(k/N - d + 1/2N)/2d} \ket{k}.
\end{equation}
Decomposing the exponential factor, we get
\begin{equation}\label{eq:uftdecom}
e^{i2\pi C} e^{-i2\pi j(d - 1/2N)/2d} e^{i2\pi jk/2dN} e^{-i2\pi k(1/2 - 1/2N)/2d},
\end{equation}
where the constant phase $C = (Nd/2 - 1/4 - 1/2d + 1/4N)/2d$. This factor will be ignored when constructing the circuit implementation of the FT, since it is eliminated when the inverse FT is applied at each step of the time-evolution operation.

Hereafter, we shall proceed mostly as proposed in Ref.~\cite{nakao} and choose the parameter $d = \frac{1}{2}$ for numerical convenience. Thus, Eq.\eqref{eq:uft1} becomes
\begin{equation}\label{eq:uft2}
U_{DFT}\ket{j} = \frac{1}{\sqrt{N}}\sum_{k=0}^{N-1} e^{-i2\pi j(1/2 - 1/2N)} e^{i2\pi jk/N} e^{-i2\pi k(1/2 - 1/2N)} \ket{k}.
\end{equation}
It is worth pointing out that if we choose a value for $d$ other than $\frac{1}{2}$, then the phases of the controlled-phase shift gates should be changed according to Eq.~\eqref{eq:uft1}. We may redefine the basis states as follows
\begin{eqnarray}\label{eq:redefbasis}
\ket{k^{\prime}} &=& e^{-i2\pi k(1/2 - 1/2N)} \ket{k}, \\
\ket{j^{\prime}} &=& e^{i2\pi j(1/2 - 1/2N)} \ket{j},
\end{eqnarray}
so that the action of $U_{DFT}$ and $U_{DFT}^\dagger$ on the new basis states gives
\begin{eqnarray}\label{eq:uft3}
U_{DFT}\ket{j^{\prime}}& =& \frac{1}{\sqrt{N}}\sum_{k=0}^{N-1} e^{i2\pi jk/N} \ket{k^{\prime}},\\
U^{\dagger}_{DFT}\ket{k^{\prime}}& = &\frac{1}{\sqrt{N}}\sum_{j=0}^{N-1} e^{-i2\pi jk/N} \ket{j^{\prime}}, 
\end{eqnarray}
in which the standard QFT are satisfied, i.e in the primed bases $U_{DFT} = U_{QFT}$. 

\subsection{Phase Estimation Algorithms}\label{sec:phase_est}
In this paper, we apply two phase estimation techniques. The quantum phase estimation algorithm is used to execute the circuits on PPRC \footnote{Plasma Physics Research Center} computer cluster. As the size of the data which could be sent to IBM servers is limited, we have to use the iterative phase estimation algorithm to reduce the size of the circuit so that it is executable on IBM devices. In this section, these two phase estimation algorithms are explained briefly. 

\subsubsection{Quantum Phase Estimation Algorithm}
In this algorithm, we use two registers; The first register contains $n$ qubits initially in the state $\ket 0$ called ancilla register. The second register begins in the state $\varphi$. So, $ \ket{\varphi_0} = \lvert 0 \rangle^{\otimes n} \lvert \varphi \rangle$. One could follow the steps below to estimate the phase:
\begin{itemize}
\item{n-bit Hadamard gate is applied to first $n$ registers:$$ \ket{\varphi_1} = {\frac {1}{2^{\frac {n}{2}}}}\left(|0\rangle +|1\rangle \right)^{\otimes n} \lvert \varphi \rangle$$.}
\item{The controlled-U gate is used so that the unitary operator $U$ is applied on the target register only if its corresponding control bit is $ \ket 1$. As
one might write $U^{2^{j}}|\varphi \rangle =e^{2\pi i2^{j}\theta }|\varphi \rangle$, applying all the $n$ controlled operations $C-U^{2^j}$ with $0\leq j\leq n-1$ gives:
\begin{eqnarray}
\ket{\varphi_{2}} & =&\frac {1}{2^{\frac {n}{2}}} \left(|0\rangle+{e^{{2\pi i} \theta 2^{n-1}}}|1\rangle \right) \otimes \cdots \otimes \left(|0\rangle+{e^{{2\pi i} \theta 2^{1}}}\vert1\rangle \right) \otimes \left(|0\rangle+{e^{{2\pi i} \theta 2^{0}}}\vert1\rangle \right) \otimes |\varphi\rangle\nonumber\\
& =& \frac{1}{2^{\frac {n}{2}}}\sum _{k=0}^{2^{n}-1}e^{{2\pi i} \theta k}|k\rangle \otimes \vert\varphi\rangle, \nonumber
\end{eqnarray}
where $k$ represents the integer representation of $n$-bit binary numbers. Note that the above expression is exactly the result of applying a QFT.}
\item{An inverse QFT is applied on ancilla register:
\begin{eqnarray}
\ket{\varphi_3} = \frac {1}{2^{\frac {n}{2}}}\sum_{k=0}^{2^{n}-1}e^{{2\pi i} \theta k}|k\rangle \otimes | \psi \rangle \xrightarrow{QFT_n^{-1}} \frac {1}{2^n}\sum_{x=0}^{2^{n}-1}\sum_{k=0}^{2^{n}-1} e^{-\frac{2\pi i k}{2^n}(x - 2^n \theta)} |x\rangle \otimes |\varphi\rangle. \nonumber
\end{eqnarray}}
\item{The above expression has a peak near $x = 2^n\theta$. If $2^n\theta$ is an integer, measuring in the computational basis gives the phase in the ancilla register with high probability:

$$ |\varphi_4\rangle = | 2^n \theta \rangle \otimes | \varphi \rangle.$$

If $2^n\theta$ is not an integer, it has been shown that the above expression has a peak about $x = 2^n\theta$ with probability better than $4/\pi^2 \approx 40\%$ \cite{qc_nielsen,qiskit_org}.}
\end{itemize}

\subsubsection{Iterative Phase Estimation Algorithm}
In this algorithm, we only use a single qubit as the work qubit register, and controlled-U operations are applied in successive stages to the system with the control bit being the work qubit. The simplest approach runs as follows \cite{O_Loan_2009}: 
\begin{itemize}
\item{Hadamard gate is applied to the work qubit in the initial state: $$\ket{\varphi_0} = \ket{0} \rightarrow \ket{\varphi_1} = \frac{1}{\sqrt{2}}(\ket{0} + \ket{1}).$$.}
\item{The controlled-U operation is applied to the system, acting on the work qubit which is very much like a phase-shift gate of the form $U_1(2\pi \theta) = \bigl( \begin{smallmatrix} 1&0\\ 0&e^{i2\pi \theta} \end{smallmatrix} \bigr)$, so that $$\ket{\varphi_1} \rightarrow \ket{\varphi_2} = \frac{1}{\sqrt{2}}(\ket{0} + e^{i2\pi \theta}\ket{1}).$$.}
\item{A final Hadamard gate on the work qubit gives the final state $$\ket{\varphi_3} = \frac{1}{2}[(1+e^{i2\pi \theta})\ket{0} + (1-e^{i2\pi \theta})\ket{1}],$$ with the probability $\frac{1}{2}[1+\cos(2\pi \theta)]$ for obtaining $\ket{0}$. A repetition of this stage is needed to obtain an estimation of $\sin(2\pi \theta)$. This can be done by applying a phase gate $S = \bigl( \begin{smallmatrix} 1&0\\ 0&i\end{smallmatrix} \bigr)$ , such that the sequence of gates is $HSU_1H$, which results in a final state  $$\ket{\varphi^\prime_3} = \frac{1}{2}[(1+ie^{2\pi \theta})\ket{0} + (1-ie^{2\pi \theta})\ket{1}],$$ with the probability $\frac{1}{2}[1-\sin(2\pi \theta)]$ for obtaining $\ket{0}$. Ultimately, with sufficient number of measurements on the work qubit, we may estimate $\sin(2\pi \theta)$ and $\cos(2\pi \theta)$ which may give us a rough estimate of $\theta$ itself.}
\end{itemize}
This algorithm is part quantum part classical. Moreover, the simulation register must be in a particular state so that the phase $\theta$, as a result of the action of the controlled-U gate on the system, can be properly estimated. In other words, we may not be able to carry out the phase estimation for a superposition of states.

\section{Circuit Analysis of the Quantum Operations}\label{sec:circop}  

In the literature of physics \cite{qc_nielsen}, the order of qubits in multi-qubit systems is usually such that the first qubit is on left most side of a tensor-product state and the last qubit is on the right. The Qiskit, however, uses a slightly different convention that if, for example, one creates a two-qubit quantum register, using the Qsikit class \texttt{QuantumRegister}, then the first qubit in the register, that is \texttt{q[0]} -where \texttt{q} is a \texttt{python list} containing the qubits- appears on the right hand side of the tensor-product state, whereas the second qubit, that is \texttt{q[1]}, appears on the left. In this convention, the left-most qubit represents the most significant bit (MSB) and the right-most qubit represents the least (LSB) \cite{qiskit_doc_summary_of_op}. This is similar to bitstring representation in classical computers, where the conversion between a bit string to its corresponding integer is as follows
 \begin{equation}\label{eq:bin2dec}
j = j_{n-1}j_{n-2} ... j_1j_0 = \sum_{m=0}^{n-1} j_m 2^{m}.
\end{equation}
This change in the representation of multi-qubit states affects the way multi-qubit gates are represented in Qiskit, and here we take this into account while constructing the quantum gates associated with the quantum operations discussed by far.

\subsection{Quantum Gate Construction of the QFT}\label{subsec:circqft}

The QFT is of significant importance and applicability in quantum computing. The QFT gate can be implemented using the Hadamard gate and controlled-phase shift gates, as shown in the following. If we pursue the instruction given in Ref.~\cite{qc_nielsen} to reach a product representation for the QFT and using the Qiskit's convention of Eq.~\eqref{eq:bin2dec}, we arrive at

\begin{eqnarray}\label{eq:qftprodrep}
\ket{j}& = &\ket{j_{n-1} ... j_0} \rightarrow \frac{1}{\sqrt{N}}\sum_{k=0}^{N-1} e^{i2\pi jk/N} \ket{k} \nonumber
\\ \nonumber\\& = & \frac{\left( \ket{0} + e^{i2\pi 0.j_0}\ket{1} \right) \left( \ket{0} + e^{i2\pi 0.j_1j_0}\ket{1}  \right) ... \left( \ket{0} + e^{i2\pi 0.j_{n-1} ... j_0} \ket{1} \right)}{\sqrt{N}}  . 
\end{eqnarray}
The inverse QFT can be simply implemented by only making the phases of the controlled-phase shift gates negative.  

\subsection{Kinetic Energy Term of the Time-evolution Operator}\label{subsec:keterm}

The quantum gate of the kinetic energy term $e^{-iK\Delta t}$ can be constructed using single-qubit phase-shift gates and controlled-phase shift gates. Note that in the momentum representation, this unitary operator becomes diagonal and very simple to implement. The implementation is instructed as follows,

\begin{equation}\label{eq:}
e^{-iK\Delta t} \ket{j} = e^{-i\frac{1}{2}p_j^2\Delta t} \ket{j}.
\end{equation}
According to Eq.~\eqref{eq:pdis}, we have
\begin{eqnarray}
p_j^2& = &(2\pi N)^2 \left(\frac{j}{N} + \frac{1}{2N} - \frac{1}{2}\right)^2 \simeq (2\pi N)^2 \left(\frac{j}{N} - \frac{1}{2}\right)^2 \nonumber
\\ \nonumber \\& = & (2\pi N)^2 \left(\frac{j^2}{N^2} - \frac{j}{N} + \frac{1}{4}\right).
\end{eqnarray}
Introducing the factor $\alpha = -(2\pi N)^2\Delta t/2$, we can rewrite
\begin{eqnarray}\label{eq:keterm1}
e^{-iK\Delta t} \ket{j}& = &e^{i\alpha (\frac{j}{N} - \frac{1}{2})^2} \ket{j} \nonumber
\\& = &\exp(i\alpha \frac{j^2}{N^2}) \exp(- i\alpha \frac{j}{N}) \exp(i\alpha\frac{1}{4}) \ket{j}, 
\end{eqnarray}
in which the last phase factor adds only a constant phase to the state of the system and hence it is sufficient to be applied on a single qubit. It is indeed important to take this constant phase factor into account as it improves the result of the phase estimation. Practically, we may only apply this phase when the state of the quantum register is known i.e. in the very beginning of the quantum circuit where the register is in state $\ket{00...0}$. 

The middle phase factor in Eq.~\eqref{eq:keterm1} depends on the particular computational basis state $\ket{j}$ it acts on, and it can be implemented using single-qubit phase-shift gates. According to Eq.~\eqref{eq:bin2dec}, one may write

\begin{equation*}
\ket{j} = \ket{j_{n-1},j_{n-2}, ... ,j_0} = \bigotimes_{m = n-1}^{0} \ket{j_m} .
\end{equation*}
Thereby,

\begin{eqnarray}\label{eq:firstterm}
e^{- i\alpha j/N}\ket{j}& = &e^{- i\alpha \sum_m j_m 2^{m-n}}\ket{j} \nonumber
\\& = & \bigotimes_{m = n-1}^{0} e^{- i\alpha j_m 2^{m-n}} \ket{j_m} \nonumber
\\& = & \bigotimes_{m = n-1}^{0} u1(- \alpha 2^{m-n}) \ket{j_m} , 
\end{eqnarray} 
where the last expression is due to the fact that $e^{- i\alpha j_m 2^{m-n}} = 1$ for $j_m = 0$ and $e^{- i\alpha j_m 2^{m-n}} = e^{- i\alpha 2^{m-n}}$ for $j_m = 1$. 

\begin{figure}
 \centering
 \begin{subfigure}{0.5\textwidth}
 \centering
 \includegraphics[keepaspectratio, width=0.95\textwidth]{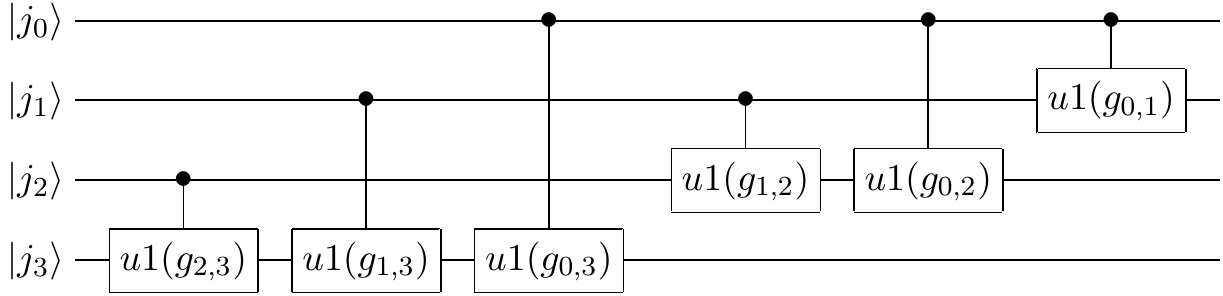}
 \caption{}
  \end{subfigure}%
 \begin{subfigure}{.5\textwidth}
 \centering
 \includegraphics[keepaspectratio, width=1\textwidth]{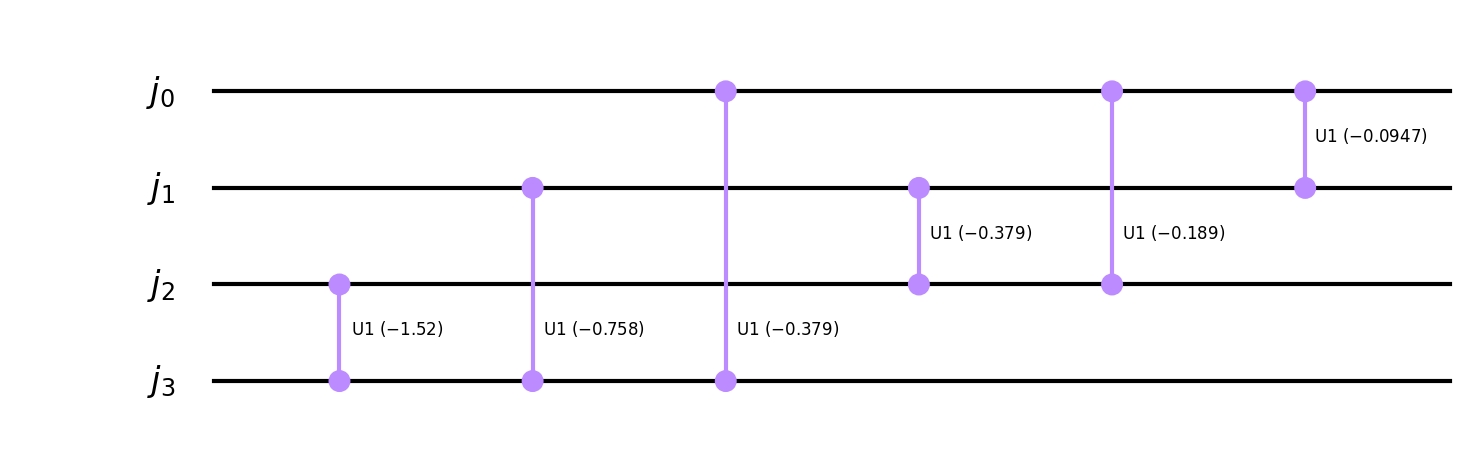}
 \caption{}
  \end{subfigure}
 \caption{ (a) Schematic circuit of $e^{i2\alpha\sum_{m<l} j_mj_l 2^{m+l-2n}}$, $g_{m,l} = 2\alpha2^{m+l-2n}$, (b) The corresponding circuit plotted in Qiskit.}
 \label{fig:circ1}
\end{figure}

The first term on the RHS of Eq.~\eqref{eq:keterm1} can be implemented using single-qubit phase-shift gates and two-qubit controlled-phase shift gates as shown below

\begin{eqnarray}\label{eq:j2n2}
\frac{j^2}{N^2}& = & (\sum_m j_m 2^{m-n})^2 \nonumber
\\& = & \sum_{m=l} j_m^2 2^{2(m-n)} + 2\sum_{m<l} j_mj_l 2^{m+l-2n}.
\end{eqnarray} 
\\
The factor $2$ in the second term of Eq.~\eqref{eq:j2n2} prevents the unnecessary repetition of controlled-phase shift gates for each pair of qubits.
Evidently, when this result is substituted back into the first term of Eq.~\eqref{eq:keterm1}, it decomposes into two exponential terms i.e. $e^{i\alpha\sum_m j_m 2^{2(m-n)}} e^{i2\alpha\sum_{m<l} j_mj_l 2^{m+l-2n}}$, where we have taken into account that $j_m^2 = j_m$. Now, the first term can be implemented by means of single-qubit phase-shift gates as in Eq.~\eqref{eq:firstterm}, i.e.

\begin{equation}
e^{i\alpha\sum_m j_m 2^{2(m-n)}} \ket{j} = \bigotimes_{m = n-1}^{0} u1(\alpha 2^{2(m-n)}) \ket{j_m} .
\end{equation}
\\
However, the second exponential term depends on the state of two qubits and hence is implemented using two-qubit controlled-phase shift gates. To understand this, consider the following equation

\begin{equation}
j_mj_l = \begin{cases}
0&      \text{if \quad $j_m = 0$ or $j_l = 0$ or $j_m = j_l = 0$ },\\
1&      \text{if \quad $j_m = j_l = 1$ },
\end{cases}
\end{equation}
\\
so that $e^{i2\alpha j_mj_l 2^{m+l-2n}}$ is an operator represented by a $4 \times 4$ matrix
 
\begin{equation*}
e^{i2\alpha j_mj_l 2^{m+l-2n}} = \begin{pmatrix}  1&0&0&0 \\ 0&1&0&0 \\ 0&0&1&0 \\ 0&0&0&e^{i2\alpha 2^{m+l-2n}} \end{pmatrix} ,
\end{equation*}
\\
which acts on each two qubits in the state $\ket{j_mj_l}$ (for $m < l$) and can be implemented using controlled-phase shift gates as shown in Fig.~\ref{fig:circ1} for the case $n=4$. In this figure, $g_{m,l} = 2\alpha2^{m+l-2n}$ and \texttt{u1} is the phase-shift gate. 

\subsection{Potential Term of the Time-evolution Operator}\label{subsec:potterm}
To simplify the implementation of the quantum algorithm, we illustrate the circuit of the potential of a particle in a finite square-well given by

\begin{equation}\label{eq:v_x}
V(x) = \begin{cases}
-V_0&      |x| < a,\\
0&            |x| > a.
\end{cases}
\end{equation}

\begin{figure}
 \centering
 \begin{subfigure}{.5\textwidth}
 \centering
 \includegraphics[width=.6\linewidth]{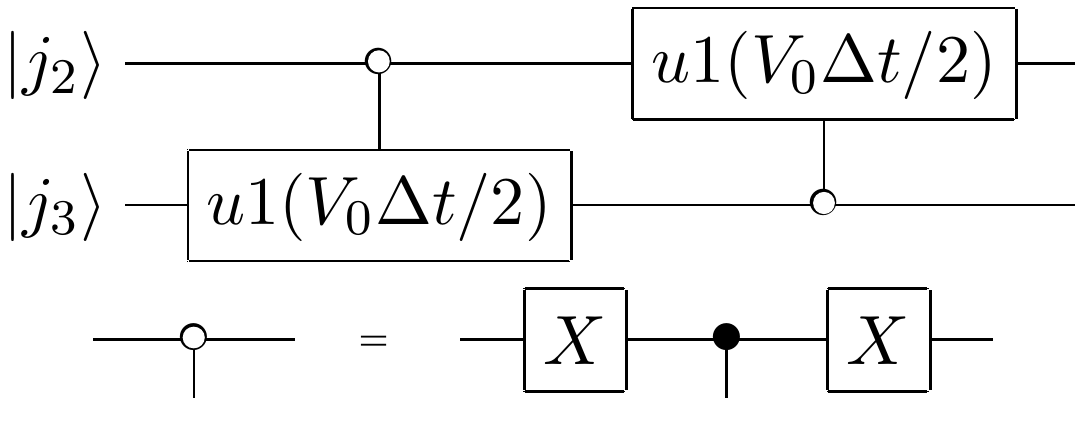}
   \caption{}
 \end{subfigure}%
 \begin{subfigure}{.5\textwidth}
 \centering
\includegraphics[width=1\linewidth]{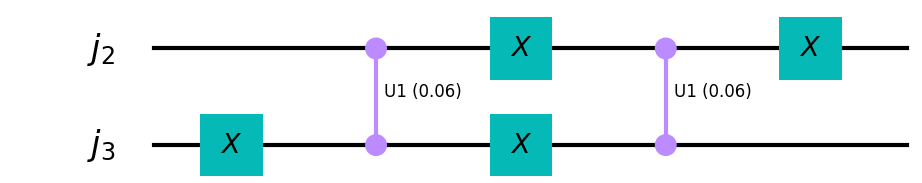}
\caption{ }
 \end{subfigure}
  \caption{(a) Schematic circuit of the potential term $e^{-iV\Delta t/2}$, (b) The corresponding circuit plotted in Qiskit; the phase of controlled-phase shift gates is determined from the values of $\Delta t$  and $V_0$.}
 \label{fig:circ2}
\end{figure}

To simulate this problem and serve its true purpose in demonstrating how adequately the quantum algorithm works, the potential strength $V_0$ and the width $a$ of the potential well should be chosen so that the particle's wave function is well localized within the region $[-\frac{1}{2},\frac{1}{2}]$. The optimized values of these parameters have been chosen to be $V_0 = 100$ and $a = \frac{1}{4}$ \cite{nakao}. We may construct the gates for the potential term $e^{-iV(x)\Delta t/2}$ rewriting Eq.~\eqref{eq:v_x} using the binary representation of the $x$-space distribution, as shown in the following
\begin{eqnarray}
|x| < a& \longrightarrow     \{x_4,x_5,x_6,x_7,x_8,x_9,x_{10},x_{11}\} ,\\
|x| > a& \longrightarrow     \{x_0,x_1,x_2,x_3,x_{12},x_{13},x_{14},x_{15}\}.
\end{eqnarray}
In the binary representation for $n = 4$, the two above sets respectively become 

\begin{align*}
\{0100,0101,0110,0111,1000,1001,1010,1011\},\\ \{0000,0001,0010,0011,1100,1101,1110,1111\}.
\end{align*}
Thereby, we can rewrite 

\begin{equation}\label{eq:v_xbin}
V(x) = \begin{cases}
-V_0&       \text{\thickspace} k_3 = 0, k_2 = 1 \text{\thickspace or \thickspace} k_3 = 1, k_2 = 0,\\
0&             \text{\thickspace} k_3 = k_2 = 0 \text{\thickspace or \thickspace} k_3 = k_2 = 1,
\end{cases}
\end{equation}
\\
where $k_m$ represents the $m$th qubit. Therefore, the potential $V(x)$ is a function of the last two qubits in our quantum register of size $n = 4$. It means that it is sufficient to implement the potential term using only two-qubit gates acting on the last two qubits (i.e. the MSB and the second one next to it), as shown in the matrix representation of the operator $e^{-iV\Delta t/2}$ below

\begin{equation*}
e^{-iV\Delta t/2} = \begin{pmatrix}  1&0&0&0 \\ 0&e^{iV_0\Delta t/2}&0&0 \\ 0&0&e^{iV_0\Delta t/2}&0 \\ 0&0&0&1 \end{pmatrix}.
\end{equation*}
\\
The implementation of this operator works most conveniently and perhaps least costly by introducing the single-qubit phase-shift gate below

\begin{equation}
u1(V_0\Delta t/2) = \begin{pmatrix}  1&0 \\ 0&e^{iV_0\Delta t/2}  \end{pmatrix}  ,
\end{equation}
which is used in the circuit shown in the Fig.~\ref{fig:circ2}. This circuit is effective and a shorter version – and thus more efficient – of the one presented in Ref.~\cite{nakao}. 

\begin{figure}
\centering
\includegraphics[width=1\linewidth]{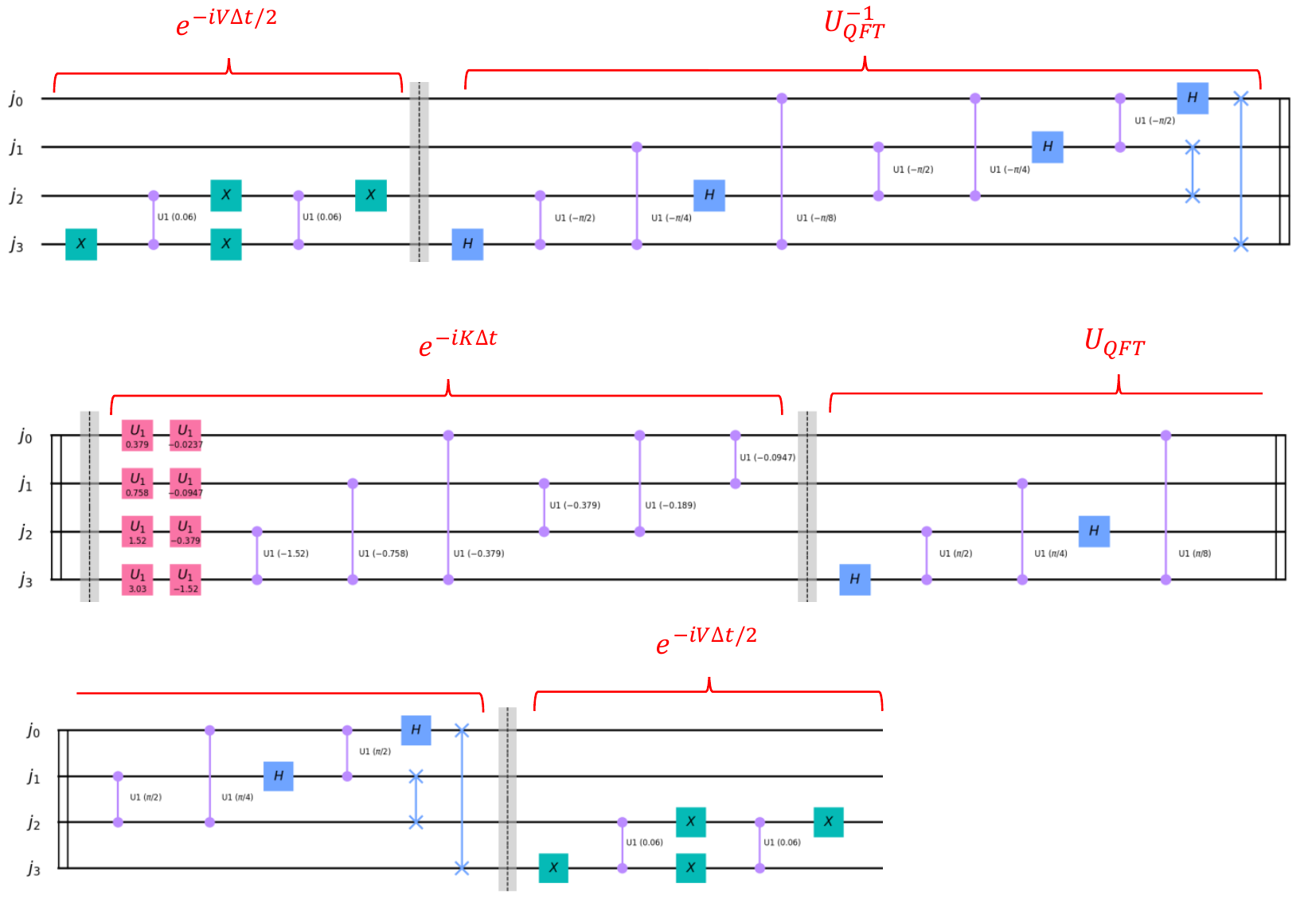}
\caption{Quantum circuit implementation of the time evolution operator on a 4-qubit register. The numbers are calculated with the chosen values for $\Delta t$ and $V_0$. If we take the work register into account, each gate is connected to a controlled qubit in the work register according to the phase estimation algorithm.} 
\label{fig:circ_time_ev}
\end{figure}

Finally, the quantum circuit corresponding to the time-evolution operator $U(\Delta t)$ implemented on a 4-qubit register, apart from the phase shift constant which only needs to be applied at the very beginning and also the controlled gates connected to the work register, is illustrated in the Fig.~\ref{fig:circ_time_ev}. 

\section{Results and Discussion}\label{sec:res}

The success of the quantum algorithm we present here tightly depends on the initial wave function of the quantum register and to which the results of the energy phase estimation are sensitive. Therefore, caution should be exercised when handling the initial preparation of the system. We use the state-initialization method suggested by Shende et al. \cite{shende} rooted in the idea that the state of the system is initially in the desired state and then it is taken back to the state $\ket{00...0}$ by constructing a quantum circuit out of CNOT and rotation gates \cite{shende}. Fortunately, QISKit comes with a built-in function \texttt{initialize}, that is easy to use, for arbitrary initialization of the quantum register, and implements the strategy just mentioned. It receives a vector of length $2^n$, as the input, representing the amplitudes $\sqrt{P_k}$ of the computational basis vectors $\ket{k}$ described in section \ref{sec:xdis}.

\begin{figure}
 \centering
 \begin{subfigure}{.5\textwidth}
 \centering
 \includegraphics[keepaspectratio, width=0.8\textwidth]{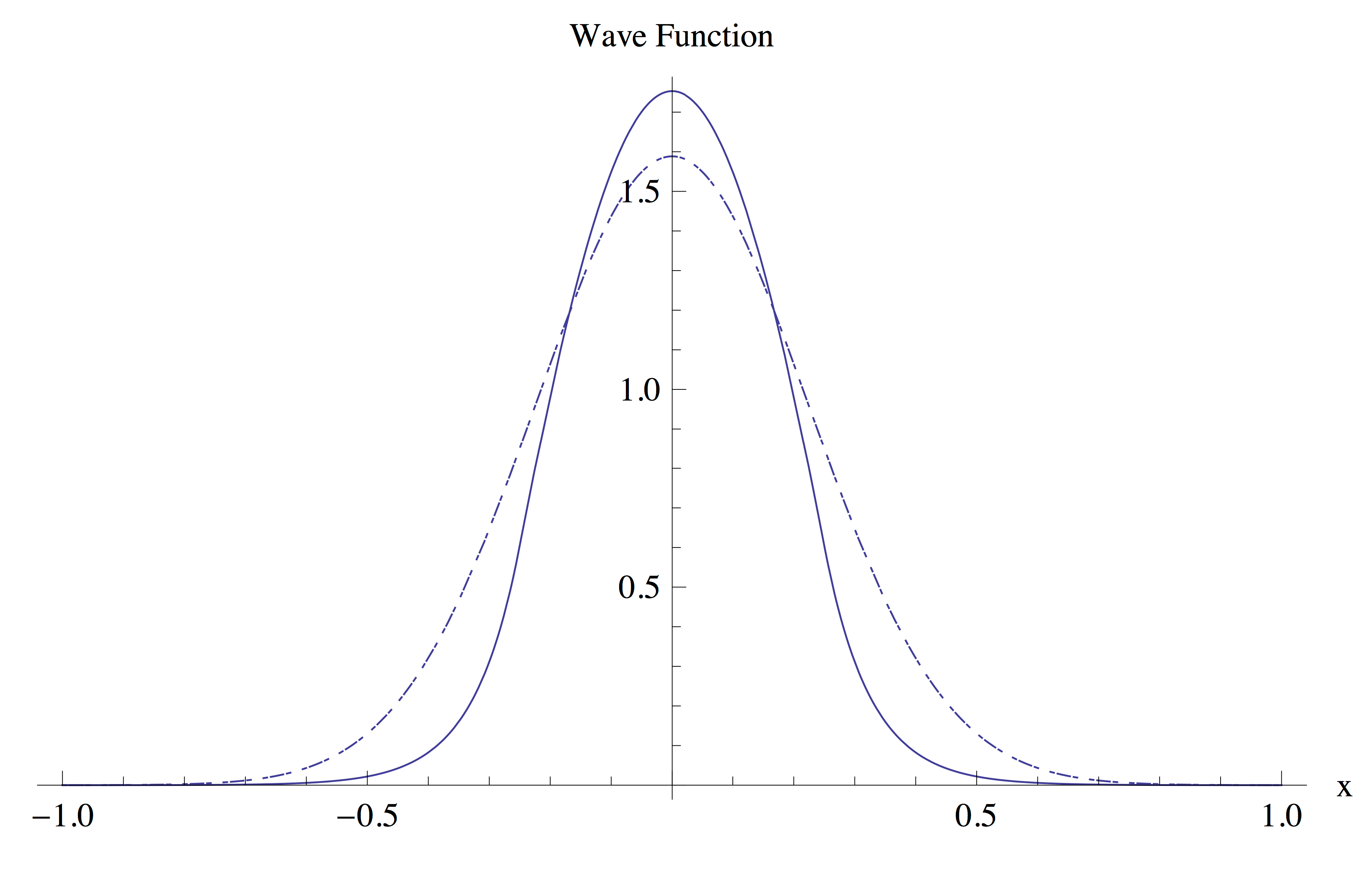}
 \caption{The ground-state wavefunction}
 \end{subfigure}%
 \begin{subfigure}{.5\textwidth}
 \centering
\includegraphics[keepaspectratio, width=1\textwidth]{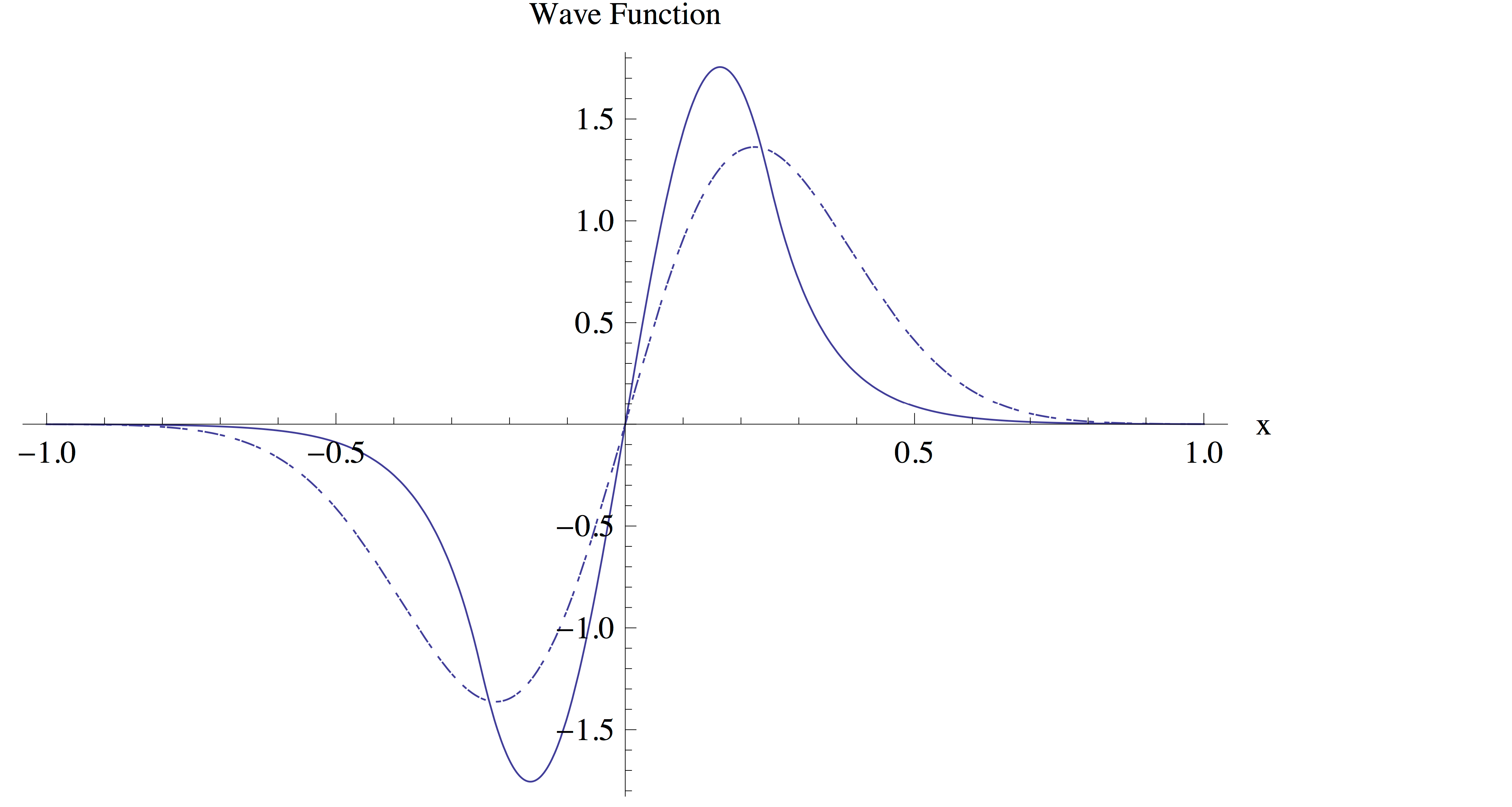}
 \caption{The first excited-state wavefunction}
 \end{subfigure}
 \caption{Comparison of the exact and trial wavefunctions in the ground-state (a) and the first excited-state (b). In each diagram, the solid line represents the exact wavefunction while the trial one is shown by the dashed line.}
 \label{fig:trial_exact_waves}
\end{figure}

Analytical eigenfunctions of the finite square-well problem corresponding to two non-degenerate energy eigenvalues are shown in Tab.~\ref{table:eig}. 
\begin{table}
\centering
\begin{tabular}{|c|c|}
\hline
\quad Eigenvalue \quad & Eigenfunction\\
\hline 
\quad$E_0 = -88.12$\quad & 

\quad$\psi_0(x) = \begin{cases}
A_0\cos(q_0x)&      |x| < a\\
C_0 e^{\pm \alpha_0 x}&            |x| > a\quad
\end{cases}$\\
\hline 
\quad$E_1 = -54.05$ \quad& 

\quad$\psi_1(x) = \begin{cases}
B_1\sin(q_1x)&      |x| < a\\
\pm C_1 e^{\pm \alpha_1 x}&            |x| > a\quad
\end{cases}$\\
\hline
\end{tabular}
\caption{Eigenvalues and eigenfunctions of the potential well. }\label{table:eig}
\text{\footnotesize $A_0 = C_0(e^{-\alpha_0a}/\cos(q_0a))$, $B_1 = -C_1(e^{-\alpha_1a}/\sin(q_1a))$, $\alpha_i^2 = 2|E_i|$, $q_i^2 = 2(100 - |E_i|)$.}
\end{table}
We recall that if one attempts to initialize the quantum register with exact eigenfunctions of the Hamiltonian, then the results of the energy phase estimation would be fairly comparable to those in Tab.~\ref{table:eig}. However, in the framework of DQS, we choose an initial state which might not precisely be the initial state of the simulated system, but some state which is a reasonable guess of the initial state. Given this point, we suggest a Gaussian wave function $N_1e^{-10 x^2}$ for the ground state and $N_2xe^{-10 x^2}$ for the first excited state with which the initial state of the register is prepared. $N_1$ and $N_2$ are the normalization constants. For a visual comparison of these trial wavefunctions with the exact ones, see Fig.~\ref{fig:trial_exact_waves}.

\begin{figure}
 \centering  
 \includegraphics[keepaspectratio, width=0.55\textwidth]{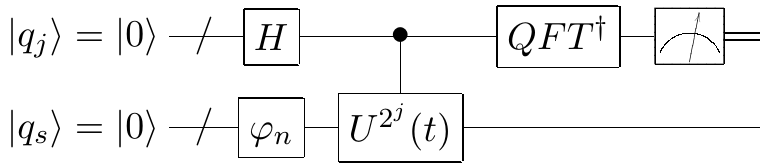} 
  \caption{Simplified version of the circuit implementation of the algorithm for estimating the energy of the one-dimensional potential-well using the quantum phase estimation executed on the PPRC computer cluster. The '/' denotes a bundle of wires. The top qubits $\ket {q_j}$ represent the work registers, where the subscript $j$ shows the qubit's index ($j \in \{0,1,...,n_w\}$), and the bottom qubits  $\ket {q_s}$ represent the simulation registers. The operation denoted by $\varphi_n$ carries out the initial preparation (which can be done by the predefined method \texttt{initialize} in QISKit) and $U^{2^j}(t)$ is the time evolution operation which has been repeated $2^j$ times.} 
\label{fig:qpe_sim_circ}
\end{figure}

To illustrate the effectiveness of the phase estimation algorithm, we demonstrate the results of a simulation carried both on a classical computer and on IBM Q's backends. We have first simulated the problem on the PPRC computer cluster\footnote{This computer includes
16 nodes and each node is equipped with two Intel Xeon X5365 CPUs. We have executed the program on the two nodes in a serial manner.} using the quantum phase estimation method. Fig.~\ref{fig:qpe_sim_circ} depicts the schematic of the overall operations involved in this simulation.

\begin{figure}
 \centering  
 \includegraphics[keepaspectratio, width=0.5\textwidth]{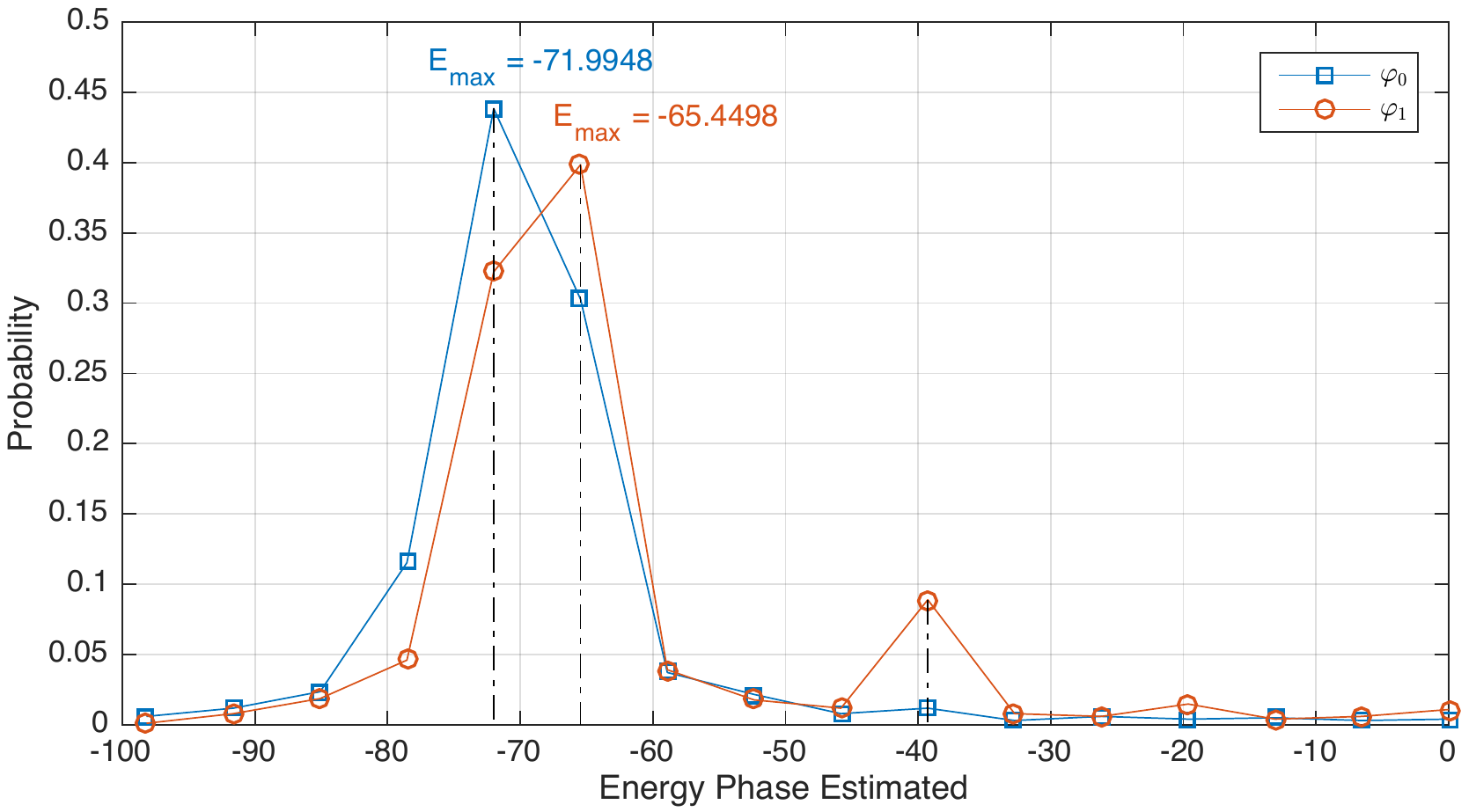} 
 \caption{Data obtained from running the algorithm on the PPRC computer cluster for the ground-state -shown by square- and the first excited state -shown by circle. In each curve, the peak indicates the most probable value of the energy phase that has been estimated by means of the quantum phase estimation algorithm after applying the time-evolution gates. The parameters have been chosen as follows:  evolution time $t = 0.06$, time step  $\Delta t = 1.2\times 10^{-3}$, total number of steps $m = 50$ and size of work and simulation registers $n_w = n_s = 4$. The initial wavefunctions are $\varphi_0(x) = N_1e^{-10x^2}$ and $\varphi_1 =  N_2 xe^{-10x^2}$, where $N_1$ and $N_2$ are normalization constants.}
\label{fig:cluster_res}
\end{figure}

Fig.~\ref{fig:cluster_res} shows the results obtained from the algorithm for the ground-state and the first excited state of the potential-well problem. The number of work and simulation registers has been chosen to be $n_w=n_s=4$. The evolution time $t$ and the time step $\Delta t$ are $0.06$ and $1.2\times 10^{-3}$, respectively with the total number of steps $m=50$. This figure basically represents the probability distribution of the qubit states (e.g. 0010 etc) of the work register after the quantum phase estimation has been carried out. 
An analysis of Fig.~\ref{fig:cluster_res} can bring out interesting points. Firstly, if we consider the most probable value of estimated energy, denoted by $E_{max}$, to be the ultimate results of the simulation, then it can be seen that the estimated energy values are somewhat close to their exact values with errors which are rooted in the limited number of qubits. Secondly, if the state $\ket{u}$ prepared on the simulation register is an eigenstate of $U$ whose eigenvalue is $e^{i2\pi \theta}$, the probability of successfully obtaining $\theta$ from the phase estimation procedure is $1-\epsilon$ with $\epsilon \ll 1$ denoting the accuracy of the phase estimation scheme. Thereby, we expect that the results of the simulation shown in Fig.~\ref{fig:cluster_res} display a peak standing on the most probable value of energy eigenvalue. In fact, this applies to the simulation associated with the first excited state, where the peak occurs for roughly the energy value $E = -72$. Tab.~\ref{table:cluster_res_err} lists the energy values obtained from the simulation along with the estimation errors as compared with the exact energy eigenvalues. 
\begin{table}
\centering
\begin{tabular}{c c c c}
State & Estimated Value & Exact Value & Errors \\
\hline\hline
Ground State & $-72.00 $ & $-88.12$ & $18.30\%$ \\
1st Excited State & $-65.44$ & $-54.05$ & $21.07\%$\\
\end{tabular}
\caption{Energy values obtained from the simulation run on the PPRC computer cluster are given in the second column with the exact energy eigenvalues in the third column. Estimation errors are shown in the last one.}\label{table:cluster_res_err}
\end{table}
In case of the ground state, apart from the large peak for the value $E = -65.45$, there is a small bump at an energy value of about $E = -39.27$, which indicates the value of the energy corresponding to the second excited state and can be explained as follows: in the preparation of the initial wave function, we have used Gaussian functions to represent the wave functions somewhat close in shape to the Hamiltonian exact eigenfunctions. Since the exact eigenfunctions form a complete set , $\{ \ket n \}$, and span the region $-d < x < d$, any square-integrable wave function which is well localized in this region of space, can be expanded by this set. Hence, it is reasonable to assume that the chosen initial wave functions, $\varphi_n$, can be expressed by an expansion of the set of Hamiltonian eigenfunctions, $\sum_m a_m \ket m$, with the probability amplitudes, $a_m$, such that $a_n$ is significantly larger than the rest, for a given $n$. In fact, we can evaluate the quantum spectrum of Hermitian matrices using the phase estimation method, when initializing the register with a superposition of eigenstates of the matrix \cite{Somma}. 

\begin{figure}
 \centering  
 \includegraphics[keepaspectratio, width=1\textwidth]{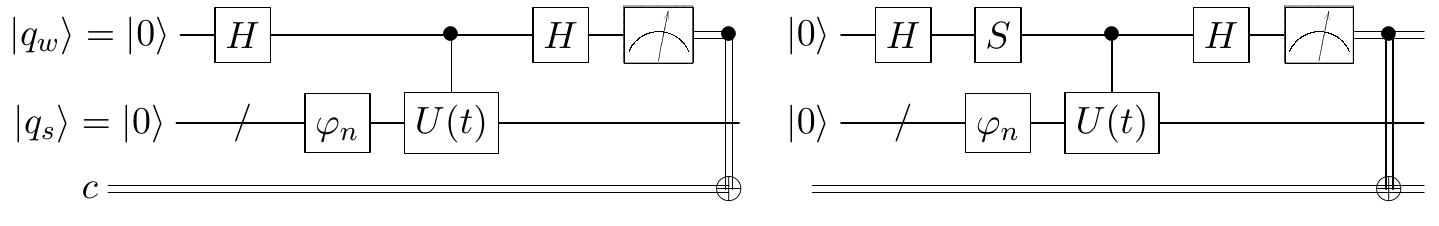} 
 \caption{The schematic of quantum circuit implementing the algorithm run on the classical backend of the IBM Q Experience. Note that in order to obtain the energy phase from the iterative phase estimation procedure two quantum  circuits needs to be executed. The LHS and RHS circuits returns $cos(-Et)$ and $sin(-Et)$ from the probability of obtaining $\ket{0}$ for the final measurement outcome, respectively.}
\label{fig:ibmq_sim_circ}
\end{figure}

Next, the classical backend IBMQ QASM simulator is used to run the algorithm. However, this time we are limited to opt for the iterative phase estimation scheme as discussed in section \ref{sec:phase_est} and this is a good chance to check the reliability of the iterative scheme. In fact, the most important constraint in this work is that the size of the data one can send to the IBM Q Experience server to execute quantum circuits -- on either the classical backend or the quantum one -- is limited, which in turn the size of the circuit to be run is limited. To work with the IBM Q Experience, one should always take into account the size of the circuits of the algorithm. This obstacle does not surface for small circuits, but for large ones, such as the one used to run the algorithm discussed in this paper. This is the primary reason why we use the iterative phase algorithm to estimate the energy. If we used a phase estimation algorithm rather than the iterative one, the controlled-$U$ gates would be applied between the simulation register and the work register successively to simulate the time evolution of the system and it is too large to be run on any of the IBM Q's backends. In terms of the circuit width in Ref.~\cite{nakao}, we use one work qubit and as usual four simulation qubits. The results do not appear as in Fig.~\ref{fig:cluster_res}, but are only numbers obtained from the iterative phase estimation algorithm. The circuit implementing this algorithm is illustrated schematically in Fig.~\ref{fig:ibmq_sim_circ}. 
Here we have used the simple approach in the iterative phase estimation technique where the time evolution gate is applied only once in each circuit. Higher number of the application of the time evolution gate over several stages results in a higher accuracy in obtaining the phase, but here we compare the results with the ones obtained from the simulation on PPRC computer cluster. So, the effort to raise accuracy is needless. The results of this simulation are shown in Tab.~\ref{table:ibmq_sim_res} for the ground state and the first excited state.
\begin{table}
\centering
\begin{tabular}{c c c c}
State & Estimated Value & Exact Value & Errors \\
\hline\hline
Ground State & $-72.56$ & $-88.12$ & $17.60\%$ \\
1st Excited State & $-65.85$ & $-54.05$ & $21.83\%$\\
\end{tabular}
\caption{The same as Tab.~\ref{table:cluster_res_err} but for the simulation done on IBMQ QASM simulator using the iterative phase estimation method.}
\label{table:ibmq_sim_res}
\end{table}
A comparison between the results shown in Tab.~\ref{table:ibmq_sim_res} and the corresponding ones shown in Tab.~\ref{table:cluster_res_err} reveals that the energy values estimated by the iterative phase estimation scheme are quite close to those obtained by means of the quantum phase estimation algorithm and hence the iterative phase estimation method works well enough. 

Finally, we do the simulation using the quantum devices of the IBM Q Experience. The first problem that one always faces when using the the IBM Q devices is the large error probability which appears when the circuit size and, especially, the number of CNOT gates increases. In fact, this outcome was observed in the form of large deviations and fluctuations of the exact values when running the circuit shown in Fig.~\ref{fig:ibmq_sim_circ}, on the quantum backends. 

Tab.~\ref{table:ibmq_qdevice_old_res} shows the results of 5 runs using the old backends IBMQX4 and IBMQX2.  
\begin{table}
\centering
\begin{tabular}{c c c c c}
Backend & $n=0$ & Error $\%$ & $n=1$ & Error$\%$ \\
\hline
\multirow{5}{4em}{IBMQX4}
& -13.42 & 84.76 & -5.30 & 93.98 \\
& -68.38 & 22.40 & -75.28 & 14.56 \\
& -76.32 & 13.38 & -43.22 & 50.95 \\
& -65.84 & 25.27 & -4.12 & 95.32 \\
& -64.43 & 26.87 & -98.17 & 11.41 \\
\rowcolor{lightgray} Mean Energy Estimated & -57.68 & 34.54 & -45.22 & 48.68 \\
\hline\hline
\multirow{5}{4em}{IBMQX2}
& -146.95 & 66.76 & -17.44 & 80.20 \\
& -78.53 & 10.87 & -86.34 & 2.01 \\
& -28.39 & 67.77 & -9.92 & 88.73 \\
& -49.30 & 44.05 & -26.17 & 70.29 \\
& -12.29 & 86.04 & -13.73 & 84.41 \\
\rowcolor{lightgray} Mean Energy Estimated & -63.09 & 28.39 & -30.72 & 65.13 \\
\hline
\end{tabular}
 \caption{The results of running the algorithm using the iterative phase estimation scheme illustrated in Fig.~\ref{fig:ibmq_sim_circ} on IBM Q old devices. The circuit for each energy level and backend has been executed 5 times. For $n = 0$, the evolution time is set to the value $t=0.04$ and the total number of steps to $m=35$, while for $n = 1$ these parameters are set to their original values; $t = 0.06$ and $m = 50$.}
 \label{table:ibmq_qdevice_old_res}
\end{table}
As for the data at hand, it can be readily seen that the numbers are scattered. This is obviously attributed to the relatively very large circuits executed in this work. Moreover, this type of simulation requires the connection between almost all qubits which leads to larger error probability. Nevertheless, looking at the results for $n = 0$ run on IBMQX4, it appears that the numbers have a lower deviation
in comparison with the data obtained from IBMQX2. This is perhaps due to the fact that we have lowered the evolution time and set it to 0.04, thus the total number of steps m has changed to 35. The modification in the parameters was necessary since the energy estimation range calculated from Eq.~\eqref{eq:E} is $E \in [0, \frac{2\pi}{t})$. The parameter $t$ was initially chosen to be $0.06$ which puts the energy in the range $[0, 100)$. This range is not large enough for the case $n = 0$ as the estimated energy of the ground state is around $E =-72$ i.e. as a consequence of scattered results, some estimated energy values go beyond this range. When this happens, the estimated values appear from the other side of the range, obviously because of the $mod$ function in the iterative phase estimation method. We
might observe a similar phenomenon in case of using the quantum phase estimation algorithm which is due to the periodic boundary condition that the QFT imposes. In simple words, if, for example, the energy value to be estimated is actually $E_{act} = -123$, then the outcomes of either of the phase estimation schemes will be around $E_{est} = -23$, that is $E_{est} = (E_{act})~mod~(\frac{2\pi}{t})$.

As the circuits executed in this work are very large and IBM Q Experience imposes limitations on the circuit size, our quantum circuits are not executable on the new backends. Given these points, we execute only the iterative phase estimation part of the circuit on the new quantum devices. In fact, the input energy values, obtained from the classical backend (see Tab.~\ref{table:ibmq_sim_res}), is given as a phase to a single simulation qubit and this phase is estimated via a work qubit using the iterative phase estimation scheme shown in Fig.~\ref{fig:ibmq_sim_circ} for the work qubit. The size of the circuits in this case has reduced to just 5 and 6. We recall that for each iterative phase estimation two circuits are required with a single $S$-gate added to the second circuit.  The results obtained from executing the circuit on the six 5 qubit devices are provided in Tab.~\ref{table:ibmq_qdevice_res} for the ground state and the first excited state. As it can be easily noticed, the scattering of the results around the exact given energy-phase values are low, which is due to the smaller size of the circuits.

\begin{table}
\centering
\begin{tabular}{c c c c c c c}
Devices & \emph{Yorktown(ibmqx2)} & \emph{Burlington} & \emph{London} &  \emph{Essex} & \emph{Vigo} & \emph{Ourense} \\
\hline\hline
Ground State & $-71.68$ & $-73.25$ & $-72.60$ & $-71.72$ & $-71.11$ & $-74.90$ \\
1st Excited State & $-68.13$ & $-57.24$ & $-65.74$ & $-62.05$ & $-53.78$ & $-68.57$ \\
\end{tabular}
\caption{Results obtained from the 5 qubit quantum devices. The evolution time is set to the value $t = 0.06$ and the total number of steps to $m = 50$.}
\label{table:ibmq_qdevice_res}
\end{table}

\section{Final Remarks}\label{sec:conclusion}

In this work, we have shown the techniques of constructing the operations and quantum gates with which the quantum circuits for the quantum simulation of a quantum particle in a simple one-dimensional potential can be created. The type of simulation investigated here is to find the energy eigenvalues by means of the phase estimation technique where an initial (trial) wave function is given to the quantum computer. As the Hamiltonian is known, the time evolution operation can be constructed by means of elementary quantum gates. Hence, the phase estimation algorithms can be used after applying the time evolution gate to find the energy spectrum of the system. Furthermore, to assess the effectiveness of the algorithm, we have used the python kit QISKit to write the code for creating the quantum circuits discussed in this work and executing them on the classical and quantum devices of the IBM Q Experience. The results obtained from executing the quantum circuits on the classical device indicate that our circuits succeed at simulating the system of the one-dimensional potential-well problem with some small errors due to the number of qubits being very small. 

A great deal of effort was paid upon reducing the size of quantum circuits in order to lower the fluctuations and large errors obtained from the results of the execution on the quantum devices, which were not very successful due to the large circuit size and large number of controlled-gates in the QFT and the phase estimation circuits. Therefore, we presented the results of the iterative phase estimation part of the whole circuit, as a part-classical part-quantum execution, so as to greatly reduce the circuit size and observe the performance of the IBM Q backends in terms of the iterative scheme.

\bibliographystyle{unsrt}
\bibliography{QC_project}

\end{document}